\documentclass[prl,superscriptaddress,showkeys,showpacs,10pt,tightenlines,floatfix,twocolumn]{revtex4-1}%
\usepackage[utf8x]{inputenc}
\usepackage[T1]{fontenc}
\usepackage{amsmath}
\usepackage{amsfonts}
\usepackage{amssymb}
\usepackage[caption=false]{subfig}
\usepackage{graphicx}
\usepackage{gensymb}
\usepackage{xcolor}
\usepackage{hyperref}
\graphicspath{{Images/}}

\newcommand{\ty}{\tilde{y}}

\graphicspath{{./images/}}

\begin{document}
	\title{Deflection and oscillations of an anchored elastic fiber embedded in a quasistatic two-dimensional foam flow}

	
	\author{Adrien Pellé and Marc Durand}
	\affiliation{Universit\'{e} Paris Cité, CNRS, UMR 7057, Mati\`{e}re et Syst\`{e}mes Complexes (MSC), F-75006 Paris, France.}
	\email{marc.durand@univ-paris-diderot.fr} 
	\date{\today}

	\begin{abstract}
We study the deflection and fluctuations of a clamped elastic fiber embedded in 2D foam under quasistatic flow.
At all times, the fiber conformation results from the elasto-capillary interactions with the foam. We independently measure the action of capillary and pressure forces on the fiber, and show that the fiber deformation is adequately described assuming a uniform continuous normal force acting on it.  When bending energy exceeds a threshold value, the fiber relaxes to a less deflected shape, generating a cascade of plastic rearrangements within the foam, and the process repeats periodically. We analyze the statistical distributions of stored and released energy, and estimate the yield stress and shear modulus of the foam, as well as the number of elementary plastic events involved in a cascade.

\end{abstract}
    \maketitle

There has been much recent interest in the coupling of a solid body with the flow of soft cellular materials, such as foams or biological tissues. These systems are constituted of highly deformable -- yet almost incompressible -- units (bubbles, drops, cells) which can slide on each other. Their multiscale composition leads to a complex rheological behavior: under small strains, they behave elastically. Above a yield value, plastic rearrangements (called T1 events) occur, conferring to these systems a complex rheological behavior \cite{Tlili_2015}. These rearrangements participate to the redistribution of stress within the system.
Even the simplest case of quasistatic-regime -- in which the structure is at mechanical equilibrium at every time and the flow is elasto-plastic-- is still being debated \cite{dollet_two-dimensional_2005,DOLLET2014731,Villemot_2021,Tlili_2020,Marmottant17271}
Analyzing the flow past undeformable obstacles allows to probe the rheological behavior for such materials  \cite{dollet_flow_2006,davies_sedimenting_2009,davies_sedimentation_2010,dollet_anti-inertial_2005,boulogne_elastoplastic_2011}. Recent studies have shown that deformable beads or droplets can be used as stress sensors  \cite{campas_quantifying_2014,Souchaud_2022}, or for probing the local mechanical properties \cite{serwane_vivo_2017}. 

The mechanical interaction of an elastic fiber with a newtonian fluid flow at low or high Reynolds number \cite{wexler_bending_2013,leclercq_drag_2016,pozrikidis_shear_2011,song_study_2021,alben_drag_2002,Buchak_2010clapping}, or with a granular flow \cite{algarra_bending_2018,seguin_buckling_2018}, received a lot of attention in past years. The challenge lies in the strong nonlinear coupling between the fluid
dynamics and the potentially large deformations of the
structure. 








In this letter we study the deflection and fluctuations of an anchored elastic fiber embedded in a two-dimensional (2D) quasistatic foam flow. 2D foam is a paradigm for other multicellular systems, in particular epithelial tissues, and is a convenient model to study the interactions with a deformable slender object, as it is easily observable, and image analysis provides information on all the geometrical properties of the foam.

We measure the respective contributions of viscous, capillary, and pressure forces on the fiber deflection, and show that the latter predominates. 
 Moreover, for the moderate fiber deflections considered here, the coupling between foam flow and fiber deflection results in a uniform pressure field along each side of the fiber, the pressure drop being concentrated
  at the fiber tip.
An independent analysis of the fiber shape also agrees with a uniform distribution of normal stress along the fiber.
 %
  We then investigate the fluctuations of the fiber deflections and determine the statistical distributions of energy stored and released by the deflected fiber, allowing us to estimate the frequency and magnitude of avalanche-like relaxation events.



\begin{figure}[h]
	\centering
	\includegraphics[width=\linewidth]{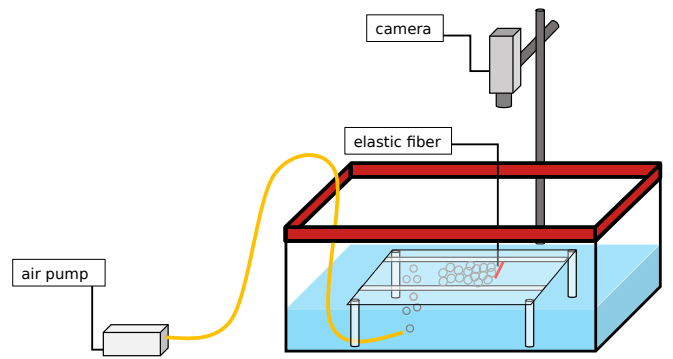}
	\caption{Sketch of the experimental setup.}
	\label{fig:setup}
\end{figure}

The experimental setup is presented on Fig. \ref{fig:setup}. 
A 20cm wide and 40cm long tank is filled with a soap solution made of a commercial dish-washing fluid 
diluted in tap water at concentration of 1.83 g/L.  The surface tension of the solution, measured with the pulling plate method, is $\gamma=25.4\pm 0.2$ mN m$^{-1}$. A gutter-shaped plate of $30$ cm length, $12$ cm width and $2$ cm depth is placed above the liquid surface with a tiny inclination, leaving a gap $h$ between the liquid surface and the covering plate that ranges from $4$ mm to $5$ mm from one side to the other of the channel length. The elastic fiber is anchored on one side of the gutter-shaped plate, at mid-length.
The cantilever we used is a soft plastic elastic beam of length $L=4.10 \pm 0.05~\mathrm{cm}$, width $w=2.5\pm 0.1~\mathrm{mm}$ and thickness $e=(140\pm 2)~\mathrm{\mu m}$. Its density is $\rho=(1.29\pm0.01)~\mathrm{kg.m^{-3}}$. Its Young modulus $E$ has been estimated using the vibrating beam method \cite{SI}, yielding $E=(2.30\pm 0.02)~\mathrm{GPa}$, a value which is compatible with typical values for common plastic materials.
A two-dimensional foam, composed of a single monolayer of bubbles confined between the liquid surface and the covering plate, is produced by blowing bubbles of air in the solution, at the entrance of the channel. The tiny slope between the cover slip and the liquid surface allows a single monolayer of bubbles to
form and flow smoothly along the channel.  The continuous gas flow is pushing forward the bubble monolayer, causing the deflection of the cantilever, until a cascade of plastic T1s occur and the fiber relaxes to a less deflected state. The foam is fairly monodisperse: distribution of bubble areas is well described with a normal law with average $0.24~\text{cm}^2$ and standard deviation = $0.08~\text{cm}^2$ \cite{SI}.
Note that at the difference with the setup used by Dollet et al. \cite{dollet_two-dimensional_2005,dollet_two-dimensional_2005b}, the obstacle (here the fiber) is entirely above the liquid surface. 
The quasistatic foam flow is recorded at cadence of 2 fps.

\begin{figure}[h]
	\centering
	\includegraphics[width=0.8\linewidth]{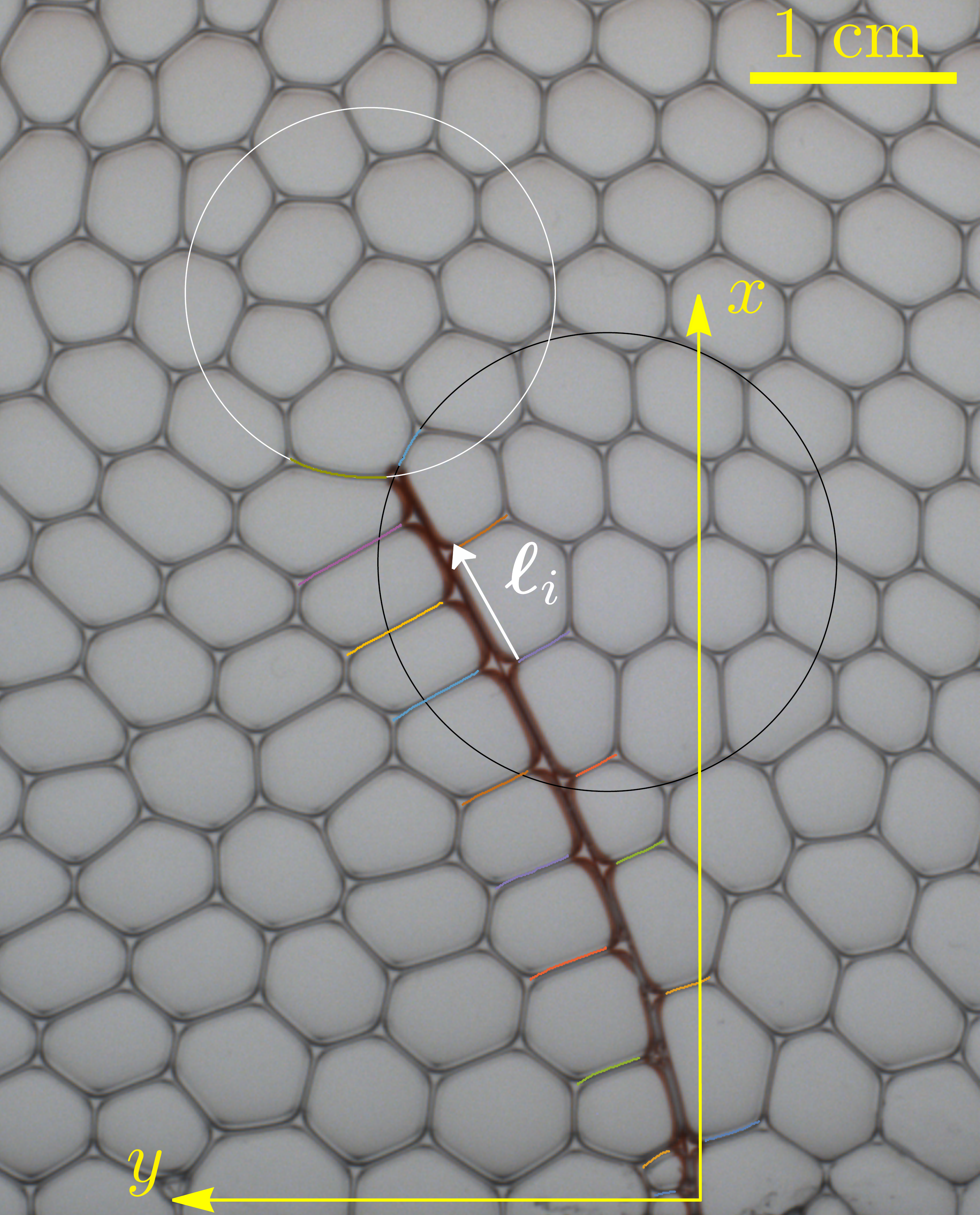}
	\caption{Snaphot of the fiber deflected by the foam flow (flowing from right to left). The corresponding number of upstream and downstream touching bubbles are, respectively, $N_u=7$ and $N_d=10$. The touching films detected by the image processing tool are colored, and the fitting circle with curvature that satisfies the criterion $\vert C_j \vert \geq \ell_j/2$ are shown (in black for the upstream side, in white for the downstream side).}
	\label{fig:fiber}
\end{figure}
A typical configuration of the deflected fiber is represented on Fig. \ref{fig:fiber}.
Before image recording, we ensured that the foam flow is quasistatic: when air blowing is stopped, the displacement of the bubble monolayer stops immediately. Moreover the bending rigidity of the fiber is low enough that it does not push the foam monolayer backward or generate structural T1 rearrangements in it to recover its undeformed shape.

Forces acting on the fiber have three origins: \textit{i)} the viscous force, \textit{ii)} the capillary force at every film that pull on both sides of the fiber, and \textit{iii)} the pressure force exerted by the gas encapsuled in every bubble in contact with the fiber. 

\textit{i) viscous force:} it has two contributions itself: the first one is associated with the sliding of bubbles along the fiber. We can estimate the associated viscous force per unit length acting tangentially on the fiber $N_f\eta w l_m (\partial v / \partial y)/L \sim N_f \eta w v/L$, where $\eta \ simeq 10^{-3} \mathrm{Pa.s}$ is the viscosity of the solution, $N_f\simeq 10$ the number of films in contact with the fiber, $v$ the mean sliding velocity of the bubbles along the fiber, and $l_m \sim 1 \text{mm}$ the typical height of a meniscus 
Actually the sliding is very limited and the same bubbles surround the fiber for most of the experiment (see videos in \cite{SI}), $V\sim 0.1 \mathrm{mm.s^{-1}}$.
 Because the fiber is not static but oscillates with time, a second contribution associated with the fiber displacements must be accounted for: films joining the fiber borders to the top plate and the pool and moving with the fiber generate a viscous force per unit length $\eta e \partial v / \partial z \sim \eta e V/h$, where $V\sim10^{-4} \mathrm{m.s^{-1}}$ is the mean fiber velocity. Both contributions yield a fiscous force per unit length $f_{vis} \lesssim 10^{-7} \mathrm{N/m}$, several orders of magnitude smaller than the other forces acting on the fiber.


\textit{ii) capillary force:} each film in contact with the fiber exerts a pulling force $\lambda=2\gamma w$ that is locally normal to it. We introduce the mean capillary force per unit length $f_\text{cap}=(N_u-N_d)\lambda/L$, where $N_u$ and $N_d$ are the upstream and downstream bubbles in contact with the fiber, and $\lambda=2\gamma w$ is the line tension of a film (the factor $2$ is here because there are two interfaces per film). Note that there is always a supplementary bubble that caps the fiber tip, so the total number of bubbles in contact is then $N_u+N_d+1$. 
Because upstream bubbles are elongated in the fiber direction whereas downstream bubbles are elongated in the flow direction, $N_u<N_d$, hence the total capillary force is oriented in the flow direction.



\textit{iii) Pressure force:} the force resulting from the pressure exerted by the touching bubbles is more difficult to evaluate. 
	In \cite{dollet_two-dimensional_2005,dollet_two-dimensional_2007} it has been estimated by analysing the area variation of the touching bubbles. The basic idea is that under in-plane compression bubbles, extend in the third direction model, resulting in a decrease of their cross-sectional area. However, this method applied to our case yields inconsistent results, as we observe that upstream bubbles have larger areas than downstream ones \cite{SI}, suggesting that the pressure force is opposed to the flow direction. Presumably, this disagreement comes from the high anisotropic deformation of bubbles which has not be accounted for in this model which assumes isotropic stress (see \cite{SI} for details). Alternatively, we estimate the pressure force from Laplace's law applied successively at every film in contact with the fiber. This method allows us to determine the bubble over-pressures (with respect to a reference bubble) with no undetermined prefactor, but requires to identify the small film curvatures with high accuracy. The resulting mean force per unit length is given by\cite{SI}:
%
\begin{equation}
	f_p =\lambda w /L \lVert  \sum_{\substack{i\in \mathcal N} } \sum_{j\leq i} C_j \boldsymbol{\ell}_j  \rVert
	\label{Force_p}
\end{equation}
where the first sum is on the set of bubbles in contact with the fiber, $C_j$ is the curvature of the film between bubble $j-1$ and $j$, and $\boldsymbol{\ell}_j $ the vector the junctions on the fiber of the two films shared by bubble $i$ (see Fig. \ref{fig:fiber}).

\begin{figure}
	\centering
	\includegraphics[width=\linewidth]{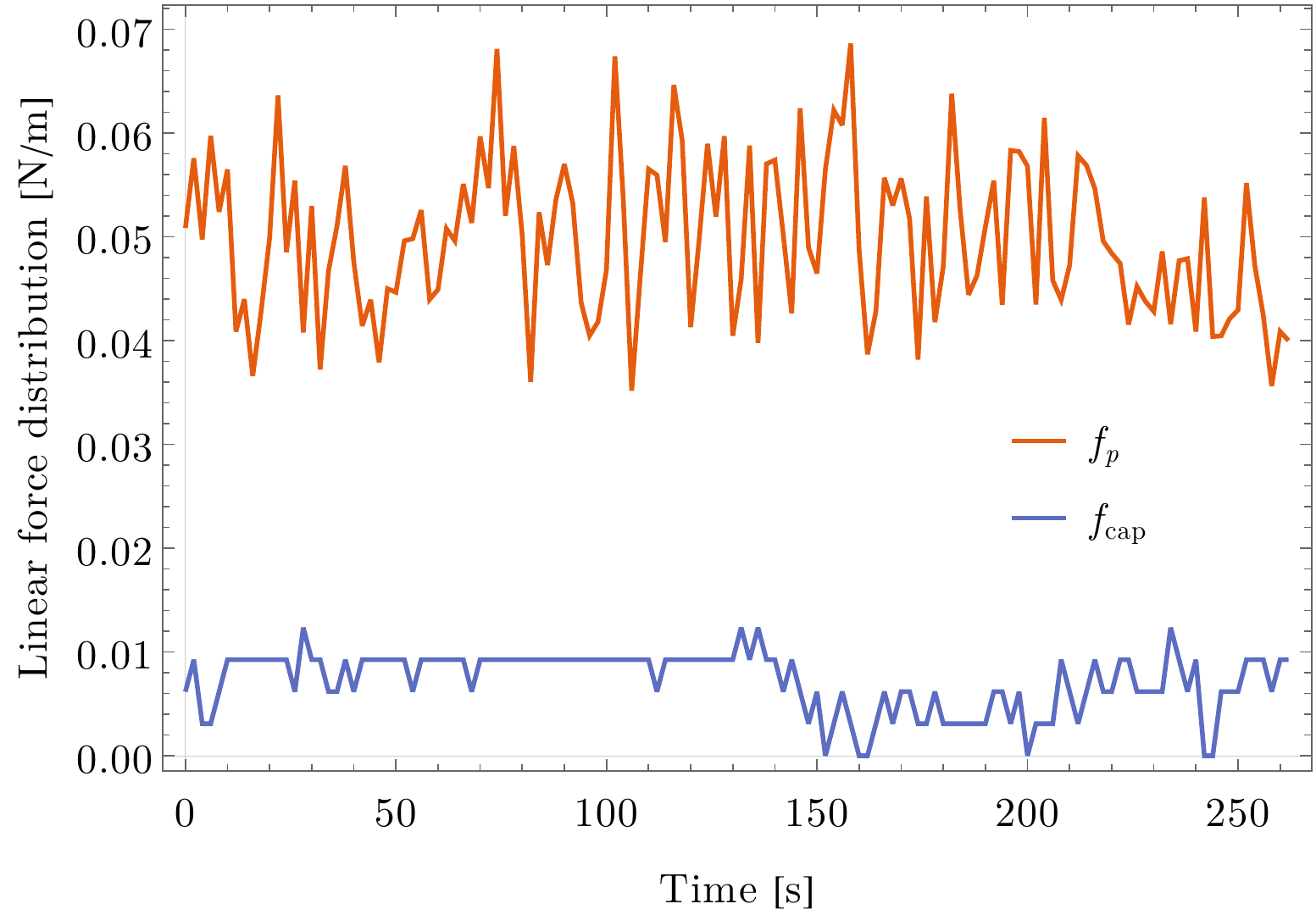}
	\caption{Typical time evolution of the capillary force $f_\text{cap}$ and pressure force $f_p$ per unit length exerted on the fiber.}
	\label{fig:correlationswcomp-wcap}
\end{figure}

Figure \ref{fig:correlationswcomp-wcap} shows $f_\text{cap}$ and $f_\text{p}$ plotted for different fiber conformations along time. The pressure force largely dominates over the capillary force, unlike what has been reported for flow past solid obstacles \cite{dollet_two-dimensional_2005}. A possible explanation for this difference is that the deformable fiber adopts conformations that reduce the difference between $N_u$ and $N_d$. 
Error bars on $f_p$ has been estimated by including more or less film curvatures in Eq. \ref{Force_p}.
Note that both acting forces are locally normal to the fiber, whatever its conformation is. Both are also much larger than the estimated viscous force.

\begin{figure}
	\centering
	\includegraphics[width=\linewidth]{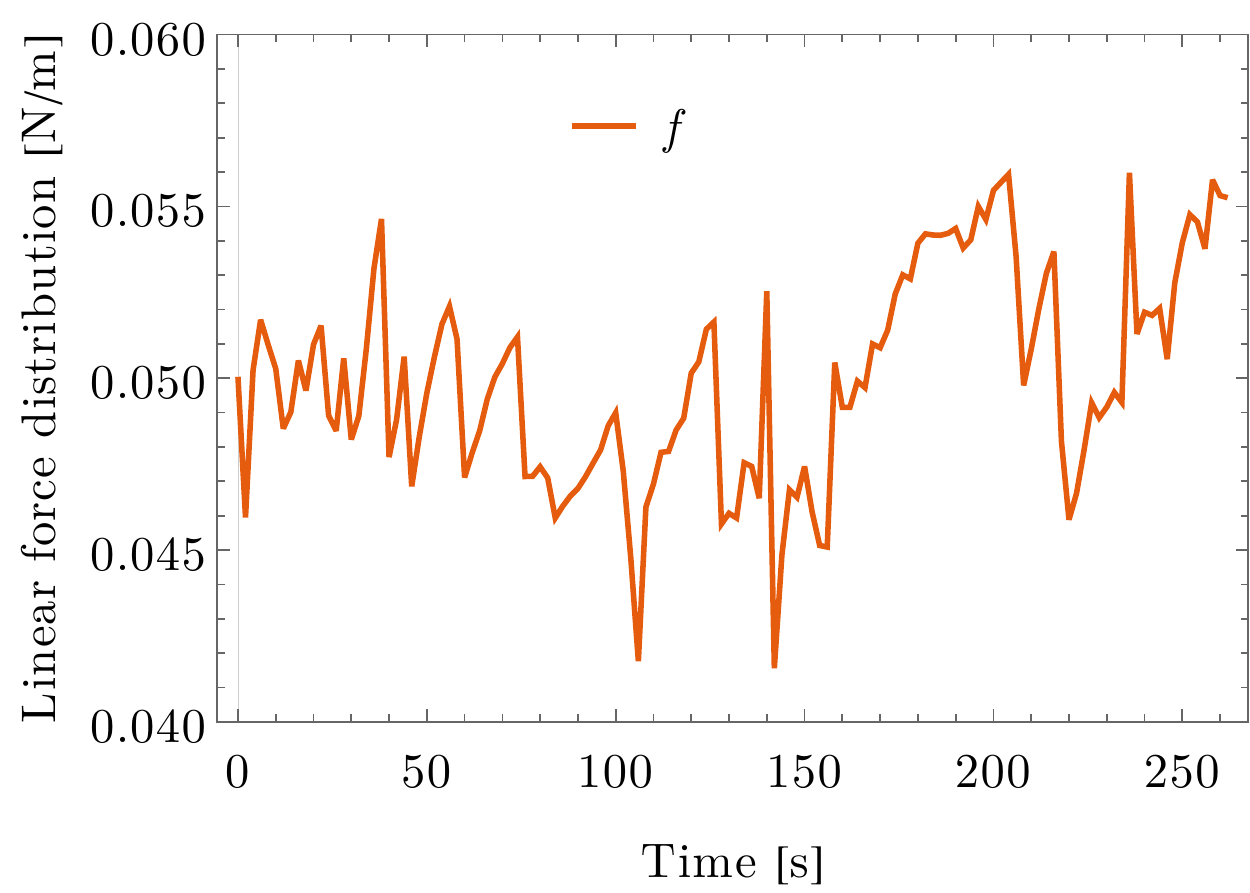}
	\caption{Typical time evolution of the linear distribution of normal force $f$ determined from the fiber deflected shape (Eq. \ref{Cartesian2}).}
	\label{fig:f_vs_time}
\end{figure}

We obtain an independent measure of the force acting on the fiber by measuring its deflected shape. Let us first note that films in contact with the fiber are nearly flat, except for the those at the vicinity of the fiber tip (see Fig. \ref{fig:fiber}), indicating that the pressure drop is mainly located there, and that pressure is uniform on each side of the fiber. Moreover, films are distributed quite evenly along the fiber. From these observations, let us assume that 
the forces acting on the fiber can be modeled as a uniform follower normal force distribution $f$. 
Because the deflection of the fiber exceeds the validity domain of the Euler-Bernoulli theory, we used a nonlinear correction to the Cartesian equation of the cantilever shape. For the moderate fiber deflections considered here, we expand the equations of mechanical equilibrium in the dimensionless parameter $\epsilon=fL^3/EI$. 
The dimensionless Cartesian equation for the fiber shape is
\begin{align}
	\ty \simeq P_1(\tilde{x}) \epsilon +P_3(\tilde{x})\epsilon^3
	\label{Cartesian2}
\end{align}
where $\tilde{x}=x/L$ $\ty=y/L$, and $P_1(\tilde{x})$ and $P_3(\tilde{x})$ are polynomials whose expressions are given in the S.I. \cite{SI}.  $P_1(\tilde{x})$ corresponds to the solution obtained in the linear regime.
We then extract $f$ from a fit of the fiber shape with the approximate solution Eq. \ref{Cartesian2}. Figure \ref{fig:f_vs_time} shows the time evolution of this normal force per unit length. Its value is comparable to $f_\text{p}$, confirming that pressure force is the main contribution to $f$.



\begin{figure}
	\centering
	\includegraphics[width=\linewidth]{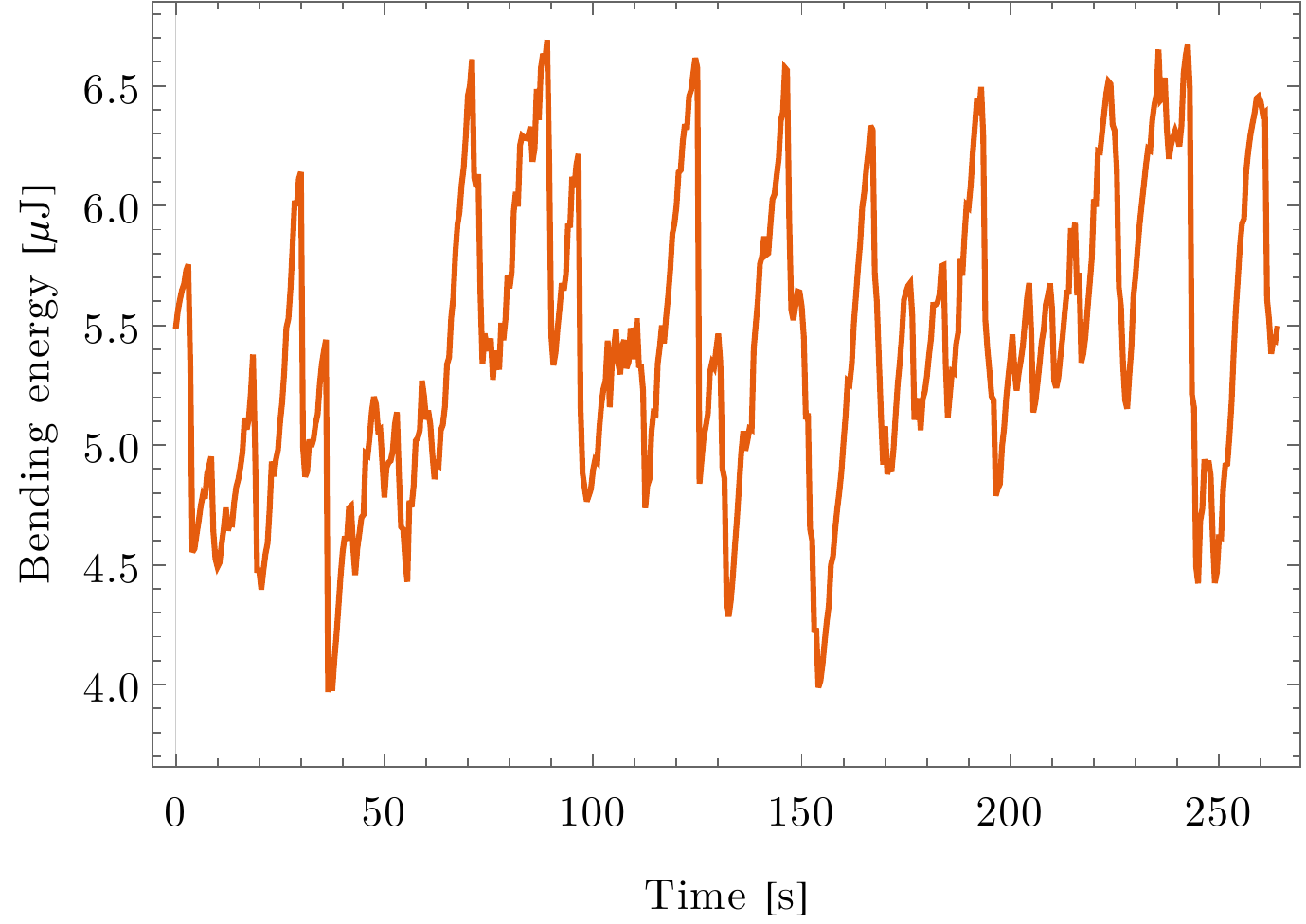}
	\caption{Typical time evolution of the fiber bending energy determined from its deflection using Eq. \ref{bending_energy}.}
	\label{fig:bendingenergyvstime}
\end{figure}
\begin{figure}
	\centering
	\includegraphics[width=\linewidth]{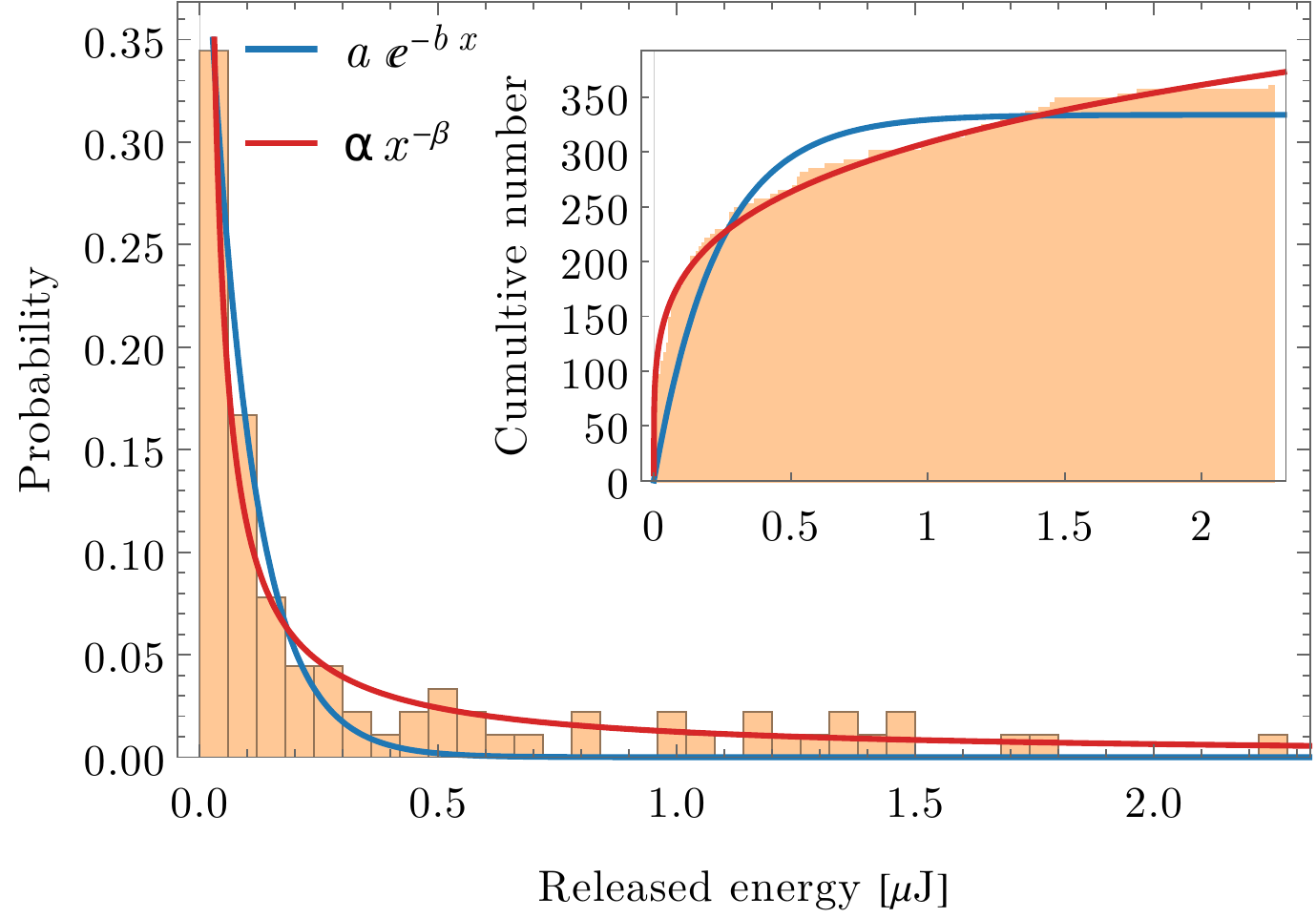}
	\caption{Histogram of energy released by the elastic fiber. Solid blue curve and solid red curve correspond to exponential $a e^{- b x}$ and power-law $\alpha x^{-\beta}$ fits, respectively. Inset: cumulative number of energy release events. Solid blue curve and solid red curve correspond to exponential $(a/b)(1- e^{- b x})$ and power-law $\alpha x^{1-\beta}/(1-\beta)$ fits, respectively.}
	\label{fig:Histogram_ReleasedEnergy}
\end{figure}
\begin{figure}
	\centering
	\includegraphics[width=\linewidth]{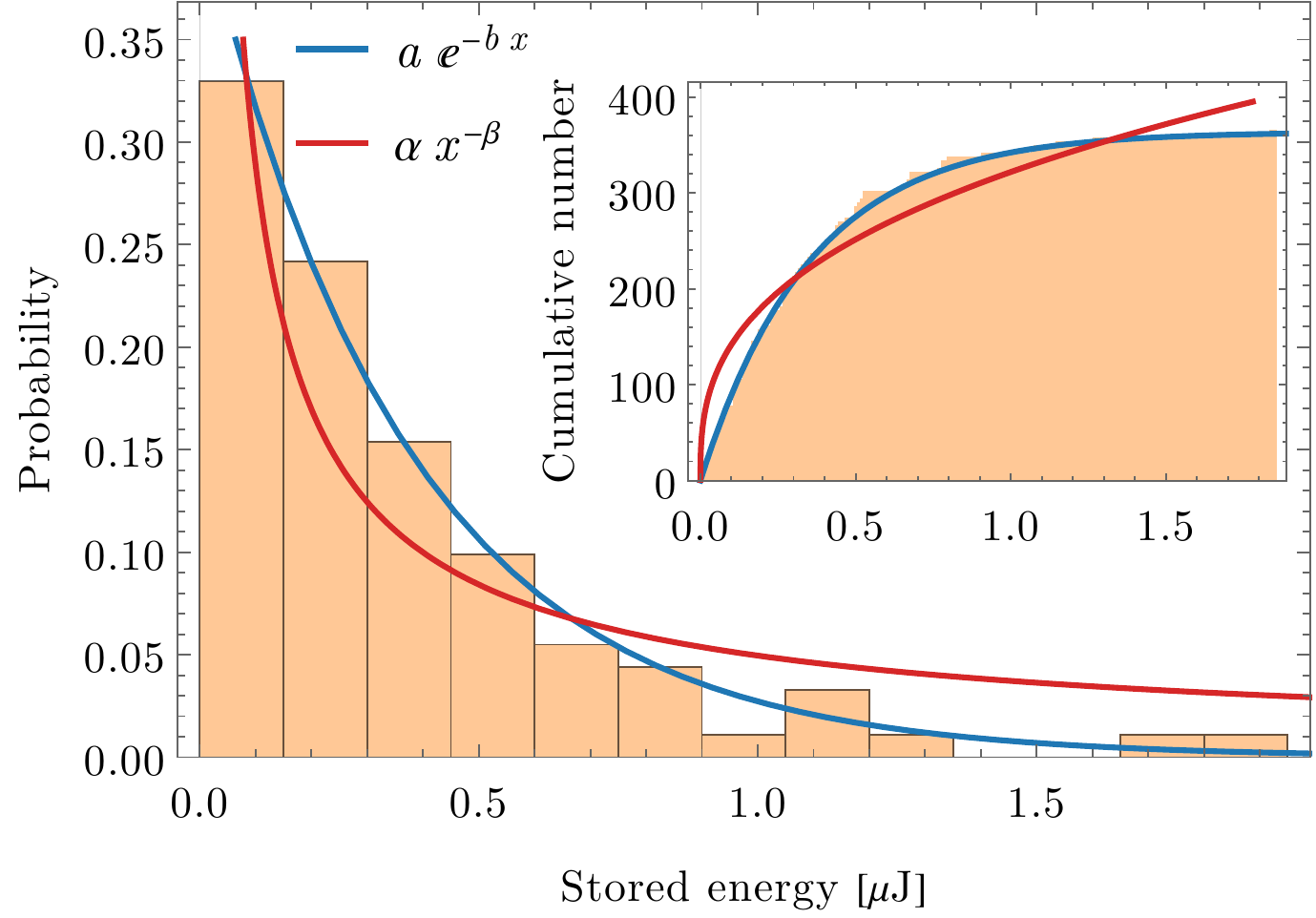}
	\caption{Histogram of energy stored by the elastic fiber. Solid blue curve and solid red curve correspond to exponential $a e^{- b x}$ and power-law $\alpha x^{-\beta}$ fits, respectively. Inset: cumulative number of energy release events. Solid blue curve and solid red curve correspond to exponential $(a/b)(1- e^{- b x})$ and power-law $\alpha x^{1-\beta}/(1-\beta)$ fits, respectively.}
	\label{fig:Histogram_StoredEnergy}
\end{figure}

We also note that the force $f$ fluctuates about its mean value $\langle f \rangle=(50\pm3)~\mathrm{mN/m}$.
The fluctuations are made more clearly visible when plotting the fiber tip deflection $\delta$, or equivalently its bending energy, given by \cite{SI}
\begin{equation}
	\mathcal F_b=\frac{EI}{L}\left(\frac{8}{5}\left(\frac{\delta}{L}\right)^2-\frac{16}{77}\left(\frac{\delta}{L}\right)^4\right).
	\label{bending_energy}
\end{equation}
Time evolution of $\mathcal F_b$ is shown in Fig. \ref{fig:bendingenergyvstime}. 
The average bending energy is $\langle \mathcal F_b \rangle=5.4~\mathrm{\mu J}$, and fluctuations are characterized by its standard deviation $\Delta \mathcal F_b=0.6~\mathrm{\mu J}$.
As shear stain is mainly located at the fiber tip, we can estimate from $f_\text{max}$ and $\delta_\text{max}$ the yield stress and shear modulus of the foam: 

The asymmetric sawtooth variations of the bending energy observed in Fig. \ref{fig:bendingenergyvstime} reveal cascades of T1 plastic events \cite{gardiner_yield_1998, Tewari_1999} that occur with quite regular periodicity (about 19s). To get further insights on the statistical properties of these cascades, we plot in Figs. \ref{fig:Histogram_ReleasedEnergy} and \ref{fig:Histogram_StoredEnergy} the histograms of released and stored energy, respectively, defined from the successive decreases and increases of the fiber bending energy.
 The average value of released energy is $\left\langle E_\mathrm{rel} \right\rangle =0.365~\mathrm{\mu J}$, with standard deviation $\Delta E_\mathrm{rel} =0.509~\mathrm{\mu J}$, 
whereas the mean stored energy is $\left\langle E_\mathrm{sto} \right\rangle =0.362~\mathrm{\mu J}$, 
with standard deviation $\Delta E_\mathrm{sto} =0.351~\mathrm{\mu J}$. The two average values are extremely close, as expected from steady regime, but the standard deviations show clear difference.

Assuming that the released bending energy of the fiber is primarily dissipated in the plastic rearrangements, its distribution gives us insight on the distribution of plastic events. In particular, it has been debated whether elementary T1 rearrangements in flowing foams occur in large cascades, or ``avalanches''  \cite{Durian_1997, Jiang_1999, Tewari_1999, Ritacco_2020}. Avalanche-like dynamics is characterized by a power-law decay of the distribution of released energy.
However, because of the limited range of energy, it is difficult to judge if the distribution of released energy is better fitted with an exponential or algebraic law (see Fig. \ref{fig:Histogram_ReleasedEnergy}). To better discriminate between an exponential and a power-law decay of the distribution, we performed fitting on the cumulative number of energy release events (see inset), which is less sensitive to the data binning. Clearly, the agreement is much better with a power-law variation, with exponent $\beta\simeq 0.77$ \cite{exponent}, revealing the avalanche-like dynamics of rearrangements.
 Note that this exponent is close to the $0.7$ value obtained numerically by Ref. \cite{Tewari_1999}.
 We also plotted the distribution of stored energy (see Fig. \ref{fig:Histogram_StoredEnergy}) and the associated cumulative number of stored energy events (inset). 
 
In contrast with the distribution of released energy, the distribution of stored energy is better adjusted with an exponential law with rate $b=2.8~\mathrm{\mu J}^{-1}$ [donne l'énergie d'un T1 unique, et donc avec mean release le nombre de T1 moyen]. We conclude that the fiber accumulates bending energy through a succession of localized deformations of the surrounding cellular material, while it relaxes it through cascades of plastic rearrangements having no specific lengthscale.





In summary, we have studied the deflection and oscillations of an elastic fiber under the quasistatic flow of a 2D foam.
We have independently measured  the different contributions to the force distribution acting along the fiber, and shown that the pressure force dominates in our configuration. We also studied the statistics on the fiber deflections and the energy released in cascades of plastic events in the surrounding foams. Finally, the measure of the maximal fiber deflection allows us to estimate the elastic modulus and the yield stress of the foam.
We hope that this study will open the way on the unexplored field of the interplay between elasto-plastic flows and deformable objects. Among possible future extensions of this work, we can cite the investigation of the large deflection regime, the deflection of a fiber aligned with the flow direction, and the interaction of an assembly of evenly disposed fibers, which represent common situations, in particular in biological systems. On the theoretical part, a continuous description of this interactions would allow to use the fiber deflection to probe the elastoplastic parameters of the foam.


\bibliographystyle{apsrev-no_url.bst}
\bibliography{biblio}

\begin{thebibliography}{31}
\expandafter\ifx\csname natexlab\endcsname\relax\def\natexlab#1{#1}\fi
\expandafter\ifx\csname bibnamefont\endcsname\relax
  \def\bibnamefont#1{#1}\fi
\expandafter\ifx\csname bibfnamefont\endcsname\relax
  \def\bibfnamefont#1{#1}\fi
\expandafter\ifx\csname citenamefont\endcsname\relax
  \def\citenamefont#1{#1}\fi
\expandafter\ifx\csname url\endcsname\relax
  \def\url#1{\texttt{#1}}\fi
\expandafter\ifx\csname urlprefix\endcsname\relax\def\urlprefix{URL }\fi
\providecommand{\bibinfo}[2]{#2}
\providecommand{\eprint}[2][]{\url{#2}}

\bibitem[{\citenamefont{{Tlili, Sham} et~al.}(2015)\citenamefont{{Tlili, Sham},
  {Gay, Cyprien}, {Graner, Fran\c{c}ois}, {Marcq, Philippe}, {Molino,
  Fran\c{c}ois}, and {Saramito, Pierre}}}]{Tlili_2015}
\bibinfo{author}{\bibnamefont{{Tlili, Sham}}},
  \bibinfo{author}{\bibnamefont{{Gay, Cyprien}}},
  \bibinfo{author}{\bibnamefont{{Graner, Fran\c{c}ois}}},
  \bibinfo{author}{\bibnamefont{{Marcq, Philippe}}},
  \bibinfo{author}{\bibnamefont{{Molino, Fran\c{c}ois}}}, \bibnamefont{and}
  \bibinfo{author}{\bibnamefont{{Saramito, Pierre}}}, \bibinfo{journal}{Eur.
  Phys. J. E} \textbf{\bibinfo{volume}{38}}, \bibinfo{pages}{33}
  (\bibinfo{year}{2015}).

\bibitem[{\citenamefont{Dollet et~al.}(2005{\natexlab{a}})\citenamefont{Dollet,
  Elias, Quilliet, Raufaste, Aubouy, and Graner}}]{dollet_two-dimensional_2005}
\bibinfo{author}{\bibfnamefont{B.}~\bibnamefont{Dollet}},
  \bibinfo{author}{\bibfnamefont{F.}~\bibnamefont{Elias}},
  \bibinfo{author}{\bibfnamefont{C.}~\bibnamefont{Quilliet}},
  \bibinfo{author}{\bibfnamefont{C.}~\bibnamefont{Raufaste}},
  \bibinfo{author}{\bibfnamefont{M.}~\bibnamefont{Aubouy}}, \bibnamefont{and}
  \bibinfo{author}{\bibfnamefont{F.}~\bibnamefont{Graner}},
  \bibinfo{journal}{Physical Review E} \textbf{\bibinfo{volume}{71}},
  \bibinfo{pages}{031403} (\bibinfo{year}{2005}{\natexlab{a}}), ISSN
  \bibinfo{issn}{1539-3755, 1550-2376}.

\bibitem[{\citenamefont{Dollet and Raufaste}(2014)}]{DOLLET2014731}
\bibinfo{author}{\bibfnamefont{B.}~\bibnamefont{Dollet}} \bibnamefont{and}
  \bibinfo{author}{\bibfnamefont{C.}~\bibnamefont{Raufaste}},
  \bibinfo{journal}{Comptes Rendus Physique} \textbf{\bibinfo{volume}{15}},
  \bibinfo{pages}{731} (\bibinfo{year}{2014}), ISSN \bibinfo{issn}{1631-0705},
  \bibinfo{note}{liquid and solid foams / Mousses liquides et solides}.

\bibitem[{\citenamefont{Villemot and Durand}(2021)}]{Villemot_2021}
\bibinfo{author}{\bibfnamefont{F.~m.~c.} \bibnamefont{Villemot}}
  \bibnamefont{and} \bibinfo{author}{\bibfnamefont{M.}~\bibnamefont{Durand}},
  \bibinfo{journal}{Phys. Rev. E} \textbf{\bibinfo{volume}{104}},
  \bibinfo{pages}{055303} (\bibinfo{year}{2021}).

\bibitem[{\citenamefont{Tlili et~al.}(2020)\citenamefont{Tlili, Durande, Gay,
  Ladoux, Graner, and Delano\"e-Ayari}}]{Tlili_2020}
\bibinfo{author}{\bibfnamefont{S.}~\bibnamefont{Tlili}},
  \bibinfo{author}{\bibfnamefont{M.}~\bibnamefont{Durande}},
  \bibinfo{author}{\bibfnamefont{C.}~\bibnamefont{Gay}},
  \bibinfo{author}{\bibfnamefont{B.}~\bibnamefont{Ladoux}},
  \bibinfo{author}{\bibfnamefont{F.}~\bibnamefont{Graner}}, \bibnamefont{and}
  \bibinfo{author}{\bibfnamefont{H.}~\bibnamefont{Delano\"e-Ayari}},
  \bibinfo{journal}{Phys. Rev. Lett.} \textbf{\bibinfo{volume}{125}},
  \bibinfo{pages}{088102} (\bibinfo{year}{2020}).

\bibitem[{\citenamefont{Marmottant et~al.}(2009)\citenamefont{Marmottant,
  Mgharbel, K{\"a}fer, Audren, Rieu, Vial, van~der Sanden, Mar{\'e}e, Graner,
  and Delano{\"e}-Ayari}}]{Marmottant17271}
\bibinfo{author}{\bibfnamefont{P.}~\bibnamefont{Marmottant}},
  \bibinfo{author}{\bibfnamefont{A.}~\bibnamefont{Mgharbel}},
  \bibinfo{author}{\bibfnamefont{J.}~\bibnamefont{K{\"a}fer}},
  \bibinfo{author}{\bibfnamefont{B.}~\bibnamefont{Audren}},
  \bibinfo{author}{\bibfnamefont{J.-P.} \bibnamefont{Rieu}},
  \bibinfo{author}{\bibfnamefont{J.-C.} \bibnamefont{Vial}},
  \bibinfo{author}{\bibfnamefont{B.}~\bibnamefont{van~der Sanden}},
  \bibinfo{author}{\bibfnamefont{A.~F.~M.} \bibnamefont{Mar{\'e}e}},
  \bibinfo{author}{\bibfnamefont{F.}~\bibnamefont{Graner}}, \bibnamefont{and}
  \bibinfo{author}{\bibfnamefont{H.}~\bibnamefont{Delano{\"e}-Ayari}},
  \bibinfo{journal}{Proceedings of the National Academy of Sciences}
  \textbf{\bibinfo{volume}{106}}, \bibinfo{pages}{17271}
  (\bibinfo{year}{2009}), ISSN \bibinfo{issn}{0027-8424},
  \eprint{https://www.pnas.org/content/106/41/17271.full.pdf}.

\bibitem[{\citenamefont{Dollet et~al.}(2006)\citenamefont{Dollet, Durth, and
  Graner}}]{dollet_flow_2006}
\bibinfo{author}{\bibfnamefont{B.}~\bibnamefont{Dollet}},
  \bibinfo{author}{\bibfnamefont{M.}~\bibnamefont{Durth}}, \bibnamefont{and}
  \bibinfo{author}{\bibfnamefont{F.}~\bibnamefont{Graner}},
  \bibinfo{journal}{Physical Review E} \textbf{\bibinfo{volume}{73}},
  \bibinfo{pages}{061404} (\bibinfo{year}{2006}), ISSN
  \bibinfo{issn}{1539-3755, 1550-2376}.

\bibitem[{\citenamefont{Davies and Cox}(2009)}]{davies_sedimenting_2009}
\bibinfo{author}{\bibfnamefont{I.}~\bibnamefont{Davies}} \bibnamefont{and}
  \bibinfo{author}{\bibfnamefont{S.}~\bibnamefont{Cox}},
  \bibinfo{journal}{Colloids and Surfaces A: Physicochemical and Engineering
  Aspects} \textbf{\bibinfo{volume}{344}}, \bibinfo{pages}{8}
  (\bibinfo{year}{2009}), ISSN \bibinfo{issn}{09277757}.

\bibitem[{\citenamefont{Davies and Cox}(2010)}]{davies_sedimentation_2010}
\bibinfo{author}{\bibfnamefont{I.}~\bibnamefont{Davies}} \bibnamefont{and}
  \bibinfo{author}{\bibfnamefont{S.}~\bibnamefont{Cox}},
  \bibinfo{journal}{Journal of Non-Newtonian Fluid Mechanics}
  \textbf{\bibinfo{volume}{165}}, \bibinfo{pages}{793} (\bibinfo{year}{2010}),
  ISSN \bibinfo{issn}{03770257}.

\bibitem[{\citenamefont{Dollet et~al.}(2005{\natexlab{b}})\citenamefont{Dollet,
  Aubouy, and Graner}}]{dollet_anti-inertial_2005}
\bibinfo{author}{\bibfnamefont{B.}~\bibnamefont{Dollet}},
  \bibinfo{author}{\bibfnamefont{M.}~\bibnamefont{Aubouy}}, \bibnamefont{and}
  \bibinfo{author}{\bibfnamefont{F.}~\bibnamefont{Graner}},
  \bibinfo{journal}{Physical Review Letters} \textbf{\bibinfo{volume}{95}},
  \bibinfo{pages}{168303} (\bibinfo{year}{2005}{\natexlab{b}}), ISSN
  \bibinfo{issn}{0031-9007, 1079-7114}.

\bibitem[{\citenamefont{Boulogne and Cox}(2011)}]{boulogne_elastoplastic_2011}
\bibinfo{author}{\bibfnamefont{F.}~\bibnamefont{Boulogne}} \bibnamefont{and}
  \bibinfo{author}{\bibfnamefont{S.~J.} \bibnamefont{Cox}},
  \bibinfo{journal}{Physical Review E} \textbf{\bibinfo{volume}{83}},
  \bibinfo{pages}{041404} (\bibinfo{year}{2011}), ISSN
  \bibinfo{issn}{1539-3755, 1550-2376}.

\bibitem[{\citenamefont{Campàs et~al.}(2014)\citenamefont{Campàs, Mammoto,
  Hasso, Sperling, O'Connell, Bischof, Maas, Weitz, Mahadevan, and
  Ingber}}]{campas_quantifying_2014}
\bibinfo{author}{\bibfnamefont{O.}~\bibnamefont{Campàs}},
  \bibinfo{author}{\bibfnamefont{T.}~\bibnamefont{Mammoto}},
  \bibinfo{author}{\bibfnamefont{S.}~\bibnamefont{Hasso}},
  \bibinfo{author}{\bibfnamefont{R.~A.} \bibnamefont{Sperling}},
  \bibinfo{author}{\bibfnamefont{D.}~\bibnamefont{O'Connell}},
  \bibinfo{author}{\bibfnamefont{A.~G.} \bibnamefont{Bischof}},
  \bibinfo{author}{\bibfnamefont{R.}~\bibnamefont{Maas}},
  \bibinfo{author}{\bibfnamefont{D.~A.} \bibnamefont{Weitz}},
  \bibinfo{author}{\bibfnamefont{L.}~\bibnamefont{Mahadevan}},
  \bibnamefont{and} \bibinfo{author}{\bibfnamefont{D.~E.}
  \bibnamefont{Ingber}}, \bibinfo{journal}{Nature Methods}
  \textbf{\bibinfo{volume}{11}}, \bibinfo{pages}{183} (\bibinfo{year}{2014}),
  ISSN \bibinfo{issn}{1548-7091, 1548-7105}.

\bibitem[{\citenamefont{Souchaud et~al.}(2022)\citenamefont{Souchaud,
  Boutillon, Charron, Asnacios, Nous, David, Graner, and
  Gallet}}]{Souchaud_2022}
\bibinfo{author}{\bibfnamefont{A.}~\bibnamefont{Souchaud}},
  \bibinfo{author}{\bibfnamefont{A.}~\bibnamefont{Boutillon}},
  \bibinfo{author}{\bibfnamefont{G.}~\bibnamefont{Charron}},
  \bibinfo{author}{\bibfnamefont{A.}~\bibnamefont{Asnacios}},
  \bibinfo{author}{\bibfnamefont{C.}~\bibnamefont{Nous}},
  \bibinfo{author}{\bibfnamefont{N.~B.} \bibnamefont{David}},
  \bibinfo{author}{\bibfnamefont{F.}~\bibnamefont{Graner}}, \bibnamefont{and}
  \bibinfo{author}{\bibfnamefont{F.}~\bibnamefont{Gallet}},
  \bibinfo{journal}{Development}  (\bibinfo{year}{2022}), ISSN
  \bibinfo{issn}{0950-1991}, \bibinfo{note}{dev.199765},
  \eprint{https://journals.biologists.com/dev/article-pdf/doi/10.1242/dev.199765/2126813/dev199765.pdf}.

\bibitem[{\citenamefont{Serwane et~al.}(2017)\citenamefont{Serwane, Mongera,
  Rowghanian, Kealhofer, Lucio, Hockenbery, and Campàs}}]{serwane_vivo_2017}
\bibinfo{author}{\bibfnamefont{F.}~\bibnamefont{Serwane}},
  \bibinfo{author}{\bibfnamefont{A.}~\bibnamefont{Mongera}},
  \bibinfo{author}{\bibfnamefont{P.}~\bibnamefont{Rowghanian}},
  \bibinfo{author}{\bibfnamefont{D.~A.} \bibnamefont{Kealhofer}},
  \bibinfo{author}{\bibfnamefont{A.~A.} \bibnamefont{Lucio}},
  \bibinfo{author}{\bibfnamefont{Z.~M.} \bibnamefont{Hockenbery}},
  \bibnamefont{and} \bibinfo{author}{\bibfnamefont{O.}~\bibnamefont{Campàs}},
  \bibinfo{journal}{Nature Methods} \textbf{\bibinfo{volume}{14}},
  \bibinfo{pages}{181} (\bibinfo{year}{2017}), ISSN \bibinfo{issn}{1548-7091,
  1548-7105}.

\bibitem[{\citenamefont{Wexler et~al.}(2013)\citenamefont{Wexler, Trinh,
  Berthet, Quennouz, du~Roure, Huppert, Lindner, and
  Stone}}]{wexler_bending_2013}
\bibinfo{author}{\bibfnamefont{J.~S.} \bibnamefont{Wexler}},
  \bibinfo{author}{\bibfnamefont{P.~H.} \bibnamefont{Trinh}},
  \bibinfo{author}{\bibfnamefont{H.}~\bibnamefont{Berthet}},
  \bibinfo{author}{\bibfnamefont{N.}~\bibnamefont{Quennouz}},
  \bibinfo{author}{\bibfnamefont{O.}~\bibnamefont{du~Roure}},
  \bibinfo{author}{\bibfnamefont{H.~E.} \bibnamefont{Huppert}},
  \bibinfo{author}{\bibfnamefont{A.}~\bibnamefont{Lindner}}, \bibnamefont{and}
  \bibinfo{author}{\bibfnamefont{H.~A.} \bibnamefont{Stone}},
  \bibinfo{journal}{Journal of Fluid Mechanics} \textbf{\bibinfo{volume}{720}},
  \bibinfo{pages}{517} (\bibinfo{year}{2013}), ISSN \bibinfo{issn}{0022-1120,
  1469-7645}.

\bibitem[{\citenamefont{Leclercq and de~Langre}(2016)}]{leclercq_drag_2016}
\bibinfo{author}{\bibfnamefont{T.}~\bibnamefont{Leclercq}} \bibnamefont{and}
  \bibinfo{author}{\bibfnamefont{E.}~\bibnamefont{de~Langre}},
  \bibinfo{journal}{Journal of Fluids and Structures}
  \textbf{\bibinfo{volume}{60}}, \bibinfo{pages}{114} (\bibinfo{year}{2016}),
  ISSN \bibinfo{issn}{08899746}.

\bibitem[{\citenamefont{Pozrikidis}(2011)}]{pozrikidis_shear_2011}
\bibinfo{author}{\bibfnamefont{C.}~\bibnamefont{Pozrikidis}},
  \bibinfo{journal}{International Journal of Solids and Structures}
  \textbf{\bibinfo{volume}{48}}, \bibinfo{pages}{137} (\bibinfo{year}{2011}),
  ISSN \bibinfo{issn}{00207683}.

\bibitem[{\citenamefont{Song et~al.}(2021)\citenamefont{Song, Chen, Yan, and
  Guo}}]{song_study_2021}
\bibinfo{author}{\bibfnamefont{J.}~\bibnamefont{Song}},
  \bibinfo{author}{\bibfnamefont{W.}~\bibnamefont{Chen}},
  \bibinfo{author}{\bibfnamefont{D.}~\bibnamefont{Yan}}, \bibnamefont{and}
  \bibinfo{author}{\bibfnamefont{S.}~\bibnamefont{Guo}}, \bibinfo{journal}{E3S
  Web of Conferences} \textbf{\bibinfo{volume}{245}}, \bibinfo{pages}{01019}
  (\bibinfo{year}{2021}), ISSN \bibinfo{issn}{2267-1242}.

\bibitem[{\citenamefont{Alben et~al.}(2002)\citenamefont{Alben, Shelley, and
  Zhang}}]{alben_drag_2002}
\bibinfo{author}{\bibfnamefont{S.}~\bibnamefont{Alben}},
  \bibinfo{author}{\bibfnamefont{M.}~\bibnamefont{Shelley}}, \bibnamefont{and}
  \bibinfo{author}{\bibfnamefont{J.}~\bibnamefont{Zhang}},
  \bibinfo{journal}{Nature} \textbf{\bibinfo{volume}{420}},
  \bibinfo{pages}{479} (\bibinfo{year}{2002}), ISSN \bibinfo{issn}{0028-0836,
  1476-4687}.

\bibitem[{\citenamefont{Buchak et~al.}(2010)\citenamefont{Buchak, Eloy, and
  Reis}}]{Buchak_2010clapping}
\bibinfo{author}{\bibfnamefont{P.}~\bibnamefont{Buchak}},
  \bibinfo{author}{\bibfnamefont{C.}~\bibnamefont{Eloy}}, \bibnamefont{and}
  \bibinfo{author}{\bibfnamefont{P.~M.} \bibnamefont{Reis}},
  \bibinfo{journal}{Physical review letters} \textbf{\bibinfo{volume}{105}},
  \bibinfo{pages}{194301} (\bibinfo{year}{2010}).

\bibitem[{\citenamefont{Algarra et~al.}(2018)\citenamefont{Algarra,
  Karagiannopoulos, Lazarus, Vandembroucq, and Kolb}}]{algarra_bending_2018}
\bibinfo{author}{\bibfnamefont{N.}~\bibnamefont{Algarra}},
  \bibinfo{author}{\bibfnamefont{P.~G.} \bibnamefont{Karagiannopoulos}},
  \bibinfo{author}{\bibfnamefont{A.}~\bibnamefont{Lazarus}},
  \bibinfo{author}{\bibfnamefont{D.}~\bibnamefont{Vandembroucq}},
  \bibnamefont{and} \bibinfo{author}{\bibfnamefont{E.}~\bibnamefont{Kolb}},
  \bibinfo{journal}{Physical Review E} \textbf{\bibinfo{volume}{97}},
  \bibinfo{pages}{022901} (\bibinfo{year}{2018}), ISSN
  \bibinfo{issn}{2470-0045, 2470-0053}.

\bibitem[{\citenamefont{Seguin and Gondret}(2018)}]{seguin_buckling_2018}
\bibinfo{author}{\bibfnamefont{A.}~\bibnamefont{Seguin}} \bibnamefont{and}
  \bibinfo{author}{\bibfnamefont{P.}~\bibnamefont{Gondret}},
  \bibinfo{journal}{Physical Review E} \textbf{\bibinfo{volume}{98}},
  \bibinfo{pages}{012906} (\bibinfo{year}{2018}), ISSN
  \bibinfo{issn}{2470-0045, 2470-0053}.

\bibitem[{SI()}]{SI}
\bibinfo{note}{See Supplemental Material [url], which includes theoretical and
  numerical details, as well as Refs. XXX}.

\bibitem[{\citenamefont{Dollet et~al.}(2005{\natexlab{c}})\citenamefont{Dollet,
  Elias, Quilliet, Huillier, Aubouy, and
  Graner}}]{dollet_two-dimensional_2005b}
\bibinfo{author}{\bibfnamefont{B.}~\bibnamefont{Dollet}},
  \bibinfo{author}{\bibfnamefont{F.}~\bibnamefont{Elias}},
  \bibinfo{author}{\bibfnamefont{C.}~\bibnamefont{Quilliet}},
  \bibinfo{author}{\bibfnamefont{A.}~\bibnamefont{Huillier}},
  \bibinfo{author}{\bibfnamefont{M.}~\bibnamefont{Aubouy}}, \bibnamefont{and}
  \bibinfo{author}{\bibfnamefont{F.}~\bibnamefont{Graner}},
  \bibinfo{journal}{Colloids and Surfaces A: Physicochemical and Engineering
  Aspects} \textbf{\bibinfo{volume}{263}}, \bibinfo{pages}{101}
  (\bibinfo{year}{2005}{\natexlab{c}}), ISSN \bibinfo{issn}{09277757}.

\bibitem[{\citenamefont{Dollet and Graner}(2007)}]{dollet_two-dimensional_2007}
\bibinfo{author}{\bibfnamefont{B.}~\bibnamefont{Dollet}} \bibnamefont{and}
  \bibinfo{author}{\bibfnamefont{F.}~\bibnamefont{Graner}},
  \bibinfo{journal}{Journal of Fluid Mechanics} \textbf{\bibinfo{volume}{585}},
  \bibinfo{pages}{181} (\bibinfo{year}{2007}), ISSN \bibinfo{issn}{0022-1120,
  1469-7645}.

\bibitem[{\citenamefont{Gardiner et~al.}(1998)\citenamefont{Gardiner,
  Dlugogorski, Jameson, and Chhabra}}]{gardiner_yield_1998}
\bibinfo{author}{\bibfnamefont{B.~S.} \bibnamefont{Gardiner}},
  \bibinfo{author}{\bibfnamefont{B.~Z.} \bibnamefont{Dlugogorski}},
  \bibinfo{author}{\bibfnamefont{G.~J.} \bibnamefont{Jameson}},
  \bibnamefont{and} \bibinfo{author}{\bibfnamefont{R.~P.}
  \bibnamefont{Chhabra}}, \bibinfo{journal}{Journal of Rheology}
  \textbf{\bibinfo{volume}{42}}, \bibinfo{pages}{1437} (\bibinfo{year}{1998}),
  ISSN \bibinfo{issn}{0148-6055, 1520-8516}.

\bibitem[{\citenamefont{Tewari et~al.}(1999)\citenamefont{Tewari, Schiemann,
  Durian, Knobler, Langer, and Liu}}]{Tewari_1999}
\bibinfo{author}{\bibfnamefont{S.}~\bibnamefont{Tewari}},
  \bibinfo{author}{\bibfnamefont{D.}~\bibnamefont{Schiemann}},
  \bibinfo{author}{\bibfnamefont{D.~J.} \bibnamefont{Durian}},
  \bibinfo{author}{\bibfnamefont{C.~M.} \bibnamefont{Knobler}},
  \bibinfo{author}{\bibfnamefont{S.~A.} \bibnamefont{Langer}},
  \bibnamefont{and} \bibinfo{author}{\bibfnamefont{A.~J.} \bibnamefont{Liu}},
  \bibinfo{journal}{Phys. Rev. E} \textbf{\bibinfo{volume}{60}},
  \bibinfo{pages}{4385} (\bibinfo{year}{1999}).

\bibitem[{\citenamefont{Durian}(1997)}]{Durian_1997}
\bibinfo{author}{\bibfnamefont{D.~J.} \bibnamefont{Durian}},
  \bibinfo{journal}{Physical Review E} \textbf{\bibinfo{volume}{55}},
  \bibinfo{pages}{1739} (\bibinfo{year}{1997}).

\bibitem[{\citenamefont{Jiang et~al.}(1999)\citenamefont{Jiang, Swart, Saxena,
  Asipauskas, and Glazier}}]{Jiang_1999}
\bibinfo{author}{\bibfnamefont{Y.}~\bibnamefont{Jiang}},
  \bibinfo{author}{\bibfnamefont{P.~J.} \bibnamefont{Swart}},
  \bibinfo{author}{\bibfnamefont{A.}~\bibnamefont{Saxena}},
  \bibinfo{author}{\bibfnamefont{M.}~\bibnamefont{Asipauskas}},
  \bibnamefont{and} \bibinfo{author}{\bibfnamefont{J.~A.}
  \bibnamefont{Glazier}}, \bibinfo{journal}{Physical Review E}
  \textbf{\bibinfo{volume}{59}}, \bibinfo{pages}{5819} (\bibinfo{year}{1999}).

\bibitem[{\citenamefont{Ritacco}(2020)}]{Ritacco_2020}
\bibinfo{author}{\bibfnamefont{H.~A.} \bibnamefont{Ritacco}},
  \bibinfo{journal}{Advances in Colloid and Interface Science}
  (\bibinfo{year}{2020}).

\bibitem[{exp()}]{exponent}
\bibinfo{note}{Note that the best fit exponent when fitting the probability
  distribution instead of the cumulative distribution is $\beta\simeq 0.95$.}

\end{thebibliography}
\end{document}